# Sparse data to structured imageset transformation

Baris Kanber[1] (ORCID: 0000-0003-2443-8800)
[1] Centre for Medical Image Computing, University College London
b.kanber@ucl.ac.uk

**Abstract.** Machine learning problems involving sparse datasets may benefit from the use of convolutional neural networks if the numbers of samples and features are very large. Such datasets are increasingly more frequently encountered in a variety of different domains. We convert such datasets to imagesets while attempting to give each image structure that is amenable for use with convolutional neural networks. Experimental results on two publicly available, sparse datasets show that the approach can boost classification performance compared to other methods, which may be attributed to the formation of visually distinguishable shapes on the resultant images.

## 1 Introduction

An important set of machine learning problems involve datasets of non-imaging origin. Such problems are usually handled effectively using conventional machine learning approaches such as support vector machines[1], or random decision forests[2]. However, as the datasets get bigger both in terms of the number of samples available, and the number of features present, these techniques can saturate. Neural networks, on the other hand, can handle large numbers of features relatively easily, and thrive when provided with large numbers of samples. Thus, they may represent a viable option for handling such datasets. One neural network architecture that is appealing for this purpose is the convolutional neural network (CNN). While it is possible to use a one-dimensional CNN in the context of sparse datasets of non-imaging origin, it may be beneficial to use convolutional network layers of two or more dimensions when the number of features is very large. One may expect to extract more performance out of the data if each sparse sample can be converted to an image (2D) or image volume (3D) that exhibits surfaces, shapes and edges that may be captured and utilised by a CNN.

A method of transforming non-image data to image format for use with CNNs was recently reported by Sharma et al. (2019). The method, authors report, constructs images with the aim of placing similar features together and dissimilar ones further apart, enabling the collective use of neighbouring elements by the CNN[3].

We take a different approach, and form images with the aim of producing visually recognizable shapes and patterns, with the hypothesis that these may be used by a CNN to improve classification performance.

## 2 Methods

We report empirical results using two datasets. The first is the Modified National Institute of Standards and Technology (MNIST) database[4] of 70000 handwritten digit images at 28x28 pixel resolution, which provides us with a sparse dataset of 70000 (number of samples) x 784 (number of features), and a 10-class classification problem. The second is the Mushroom dataset from the University of California at Irvine Machine Learning Repository[5], 8214 instances of mushroom species described in terms of 22 physical characteristics, and a target variable which identifies the specifies as edible or poisonous. The features are subsequently encoded, resulting in a sparse dataset of 8214 (number of samples) x 119 (number of features), and a binary classification problem.

We write a sparse data matrix of shape M (number of samples) x N (number of features) as D and convert it to an image matrix I of shape M x P x P where P is given by the smallest even number that is greater than or equal to $N^{1/2}$.

Five transformation schemes are explored. The conversion scheme ASIS transforms the sparse data matrix D to an imageset I using a linear filling order (Figure 1), keeping the ordering of the sparse features intact. In other words, for the M x 8 x 8 imageset depicted in Figure 1, we set: I[:,0,0]=D[:,0], I[:,0,1]=D[:,1], I[:,0,2]=D[:,2], and so on.

**Figure 1** - The linear filling order for an M x 8 x 8 imageset. This implies vector $I_{0..M-1,0,0}$ is populated first, followed by $I_{0..M-1,0,1}$, $I_{0..M-1,0,2}$, and so on.

| x  |    |    |    |    |    |    |    |
|----|----|----|----|----|----|----|----|
| y  |    |    |    |    |    |    |    |
| 0  | 1  | 2  | 3  | 4  | 5  | 6  | 7  |
| 15 | 14 | 13 | 12 | 11 | 10 | 9  | 8  |
| 16 | 17 | 18 | 19 | 20 | 21 | 22 | 23 |
| 31 | 30 | 29 | 28 | 27 | 26 | 25 | 24 |
| 32 | 33 | 34 | 35 | 36 | 37 | 38 | 39 |
| 47 | 46 | 45 | 44 | 43 | 42 | 41 | 40 |
| 48 | 49 | 50 | 51 | 52 | 53 | 54 | 55 |
| 63 | 62 | 61 | 60 | 59 | 58 | 57 | 56 |

The conversion scheme RAND also uses a linear filling order but randomises the ordering of the sparse features, while the conversion scheme DI uses the method of Sharma et al. (2019). The transformation scheme SDIC uses the linear filling regime akin to ASIS, and RAND but transforms sparse dataset $D_{MN}$ to imageset $I_{MPP}$ ordering the features according to the Pearson product-moment correlation coefficients (CORRMATRIX) of the sparse features.

Briefly, we find the largests element $CORRMATRIX_{ij}$ of CORRMATRIX and select `i` and `j` as the seed features. These are used as the first, and second features of the transformation, and they, along with any subsequently used features are marked not to be selected a second time. The next feature selected is the feature `k` for which the element $CORRMATRIX_{jk}$ of CORRMATRIX is the largest. The latter process is repeated until all features have been used, except for when $CORRMATRIX_{jk}$ is less than or equal to 0, in which case two new seeding features are found, and the process continued.

The fifth, and final transformation scheme $SDIC_C$ uses the same method as the transformation scheme SDIC except for the use of a circular filling order (Figure 2). This means, for the M x 8 x 8 imageset depicted in Figure 1, we set: I[:,3,3]=D[:,F$_1$], I[:,2,2]=D[:,F$_2$], I[:,3,2]=D[:,F$_3$], and so on, where F$_n$ is the n$^{th}$ selected feature.

**Figure 2** - The circular filling order. This implies vector I$_{0..M-1,3,3}$ is populated first, followed by I$_{0..M-1,2,2}$, I$_{0..M-1,3,2}$, and so on.

| x →    |    |    |    |    |    |    |    |
|----|----|----|----|----|----|----|----|
| 26 | 49 | 48 | 47 | 46 | 45 | 44 | 64 |
| 27 | 10 | 25 | 24 | 23 | 22 | 43 | 63 |
| 28 | 11 | 2  | 9  | 8  | 21 | 42 | 62 |
| 29 | 12 | 3  | 1  | 7  | 20 | 41 | 61 |
| 30 | 13 | 4  | 5  | 6  | 19 | 40 | 60 |
| 31 | 14 | 15 | 16 | 17 | 18 | 39 | 59 |
| 32 | 33 | 34 | 35 | 36 | 37 | 38 | 58 |
| 50 | 51 | 52 | 53 | 54 | 55 | 56 | 57 |

(y ↓)

In the case of the DI, SDIC, and SDIC$_C$ transformation schemes, fitting is done on training data, and the same transformation is applied to validation, and test data. Classification performance of each method is evaluated using bootstrapping whereby we randomly allocate the available data to training, validation, and testing sets during each bootstrap. The different transformation methods are evaluated using the same data during each bootstrap. For the MNIST dataset, the allocation is in the ratio 60000:5000:5000 for training, validation, and testing, respectively. For the MUSHROOM dataset, the same ratio is 4124:2000:2000. Additionally, to increase the complexity of the latter, we invert 20 of the 119 binary sparse features during each bootstrap.

The evaluation is performed using a CNN of architecture CONV2D->CONV2D->MAXPOOLING->DROPOUT->DENSE->DROPOUT->DENSE. The first two layers are 2D convolutional layers for which the numbers of filters, and the filter sizes are 32 and 64, and 3x3 for the evaluation of the MNIST dataset, while they are 36 and 72, and 6x6 for the MUSHROOM dataset. The maxpooling layer is 2x2 in both cases. The dropout probabilities are 0.25 and 0.5 for the MNIST evaluation, while they are 0.3, and 0.2 in the MUSHROOM dataset case. The dense layers have 128 and 10 neurons for MNIST, while they are 64, and 2 for MUSHROOM. The activation function for the first dense layer is RELU, and for the second layer it is SOFTMAX, in both cases. We use cross entropy as the loss function and Adadelta as the optimiser[6,7].

We additionally evaluate the performance of the different transformation schemes using a Random Forest decision tree classifier (RF)[8,9], which has 2000 estimators in evaluation of both the MNIST, and the MUSHROOM datasets.

We use the classification accuracy and the Area under the Receiver Operating Characteristic Curve (AUC) as the performance metrics. In the case of the 10-class MNIST classification problem, compute the average AUC of all possible pairwise combinations of classes, which is reported to be insensitive to class imbalance[10].

## 3    Results

Examples of transformations from the MNIST dataset are shown in Figure 3. On the MNIST dataset, the SDIC transformation scheme achieved the highest performance bar ASIS (Figure 4). The latter, in this case were the original MNIST handwritten digit images. This was followed by $SDIC_C$, RAND and DI transformation schemes, while RF had the lowest performance. The results were concordant with the corresponding classification accuracies, except for DI, which performed better than RAND (Table 1).

**Figure 3** - Examples of transformations from the MNIST dataset.

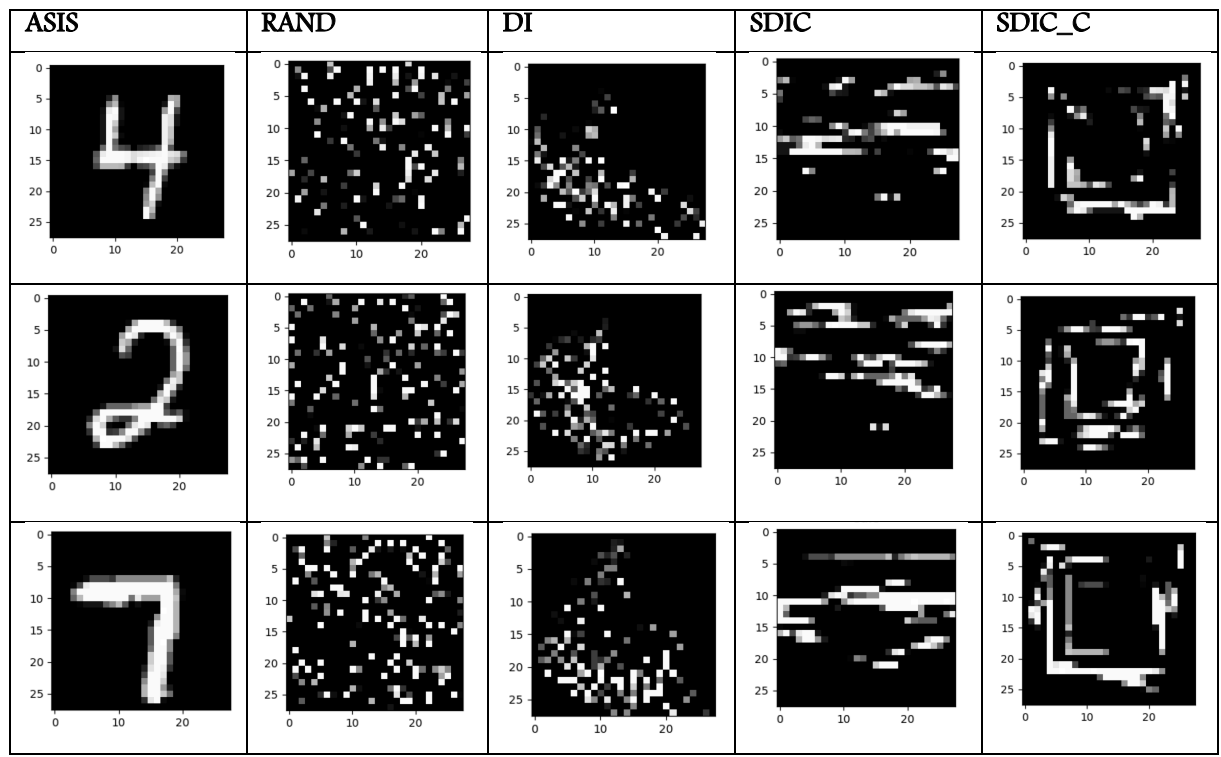

**Table 1** – Classification accuracies of the different transformation schemes on the MNIST dataset.

| Method | Test accuracy, median [interquartile range] |
|---|---|
| ASIS | 0.992 [0.991-0.993] |
| RAND | 0.978 [0.975-0.980] |
| SDIC | 0.984 [0.983-0.985] |
| $SDIC_C$ | 0.983 [0.982-0.984] |
| DI | 0.980 [0.982-0.978] |
| RF | 0.969 [0.967-0.973] |

**Figure 4** – Performance (AUC) of the different transformation schemes on the MNIST dataset.

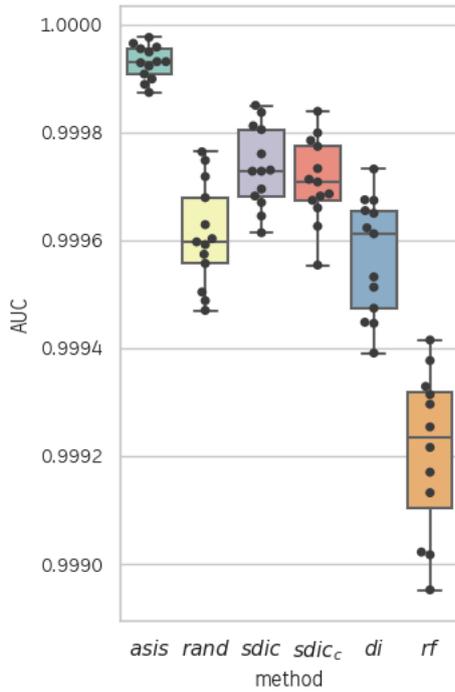

Examples of transformations from the MUSHROOM dataset are shown in Figure 5. On this dataset, the transformation scheme SDIC$_C$ had the highest performance, followed by RF (Figure 6). The transformation scheme ASIS performed somewhat similar to, but poorer than RF, while RAND, SDIC, and DI had sub-optimal performance. These results were consistent with the corresponding classification accuracies (Table 2).

**Figure 5** - Examples of transformations from the MUSHROOM dataset.

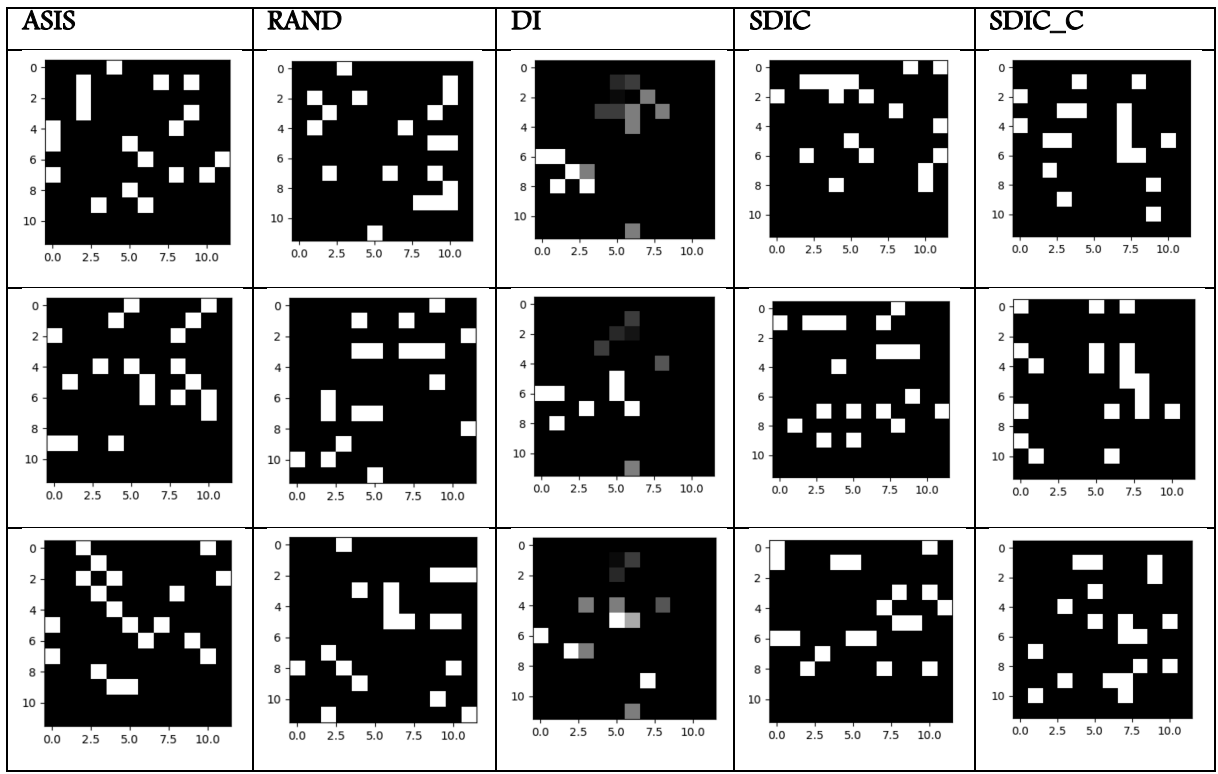

Table 2 – Classification accuracies of the different transformation schemes on the MUSHROOM dataset.

| Method | Test accuracy, median [interquartile range] |
|---|---|
| ASIS | 0.950 [0.949-0.954] |
| RAND | 0.948 [0.940-0.953] |
| SDIC | 0.945 [0.944-0.945] |
| $SDIC_C$ | 0.958 [0.955-0.961] |
| DI | 0.940 [0.936-0.944] |
| RF | 0.956 [0.949-0.958] |

Figure 6 – Performance (AUC) of the different transformation schemes on the MUSHROOM dataset.

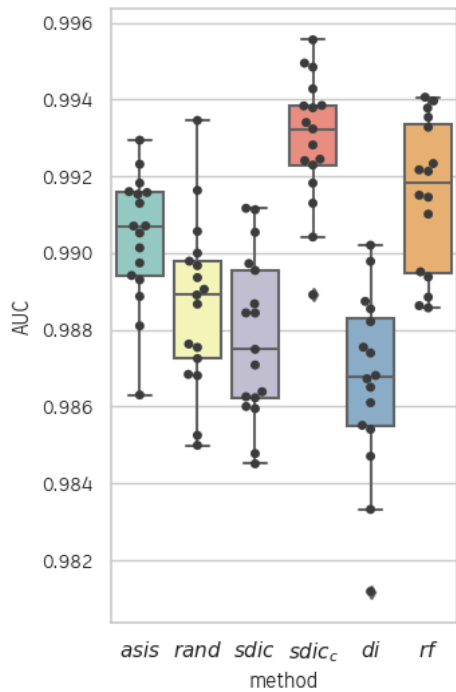

## 4   Discussion

In the case of MNIST, the ASIS transformation scheme was a null transformation as this reconstructs the original handwritten digit images. Performance of this transformation scheme on the MNIST dataset is therefore the performance of the network on the original MNIST dataset itself. In this context, performance of the null transformation ASIS represents what an efficient transformation method should aim to approach. On this dataset, SDIC was the best performing non-null transformation, closely followed by $SDIC_C$ (Figure 4). Of note was the relatively good performance of the random transformation RAND, which suggested that the distribution of the features across the field-of-view of the image may have some benefits. The RAND transformation, also performed better than RF on the MNIST dataset. This is simply a manifestation of MNIST being a classification problem of imaging origin, and as such a traditional approach such as the random forest decision tree classifier, lags behind a CNN operating on a visually uninterpretable version of the handwritten digits (Figure 3).

On the MUSHROOM dataset, SDIC$_C$ was the best performing transformation method, closely followed by RF (Figure 6). We find that, in this problem of non-imaging origin, it is difficult to surpass the performance of a traditional classifier such as RF, using a CNN. However, the MUSHROOM dataset contained a relatively low number of features, preventing the formation of discerable patterns on the resultant images (Figure 5). The DI transformation method performed poorly in this case, but this may be attributed to the lossy compression of the resultant images to a size smaller than is required by the technique[3].

Further work can explore whether these techniques bring benefit in a diverse range of classification/regression problems involving large, sparse datasets. It should also be investigated whether the feature ordering schemes used in the SDIC and SDIC$_C$ transformation methods could benefit from the addition of a prior, unsupervised, clustering stage. For example, in the case of the MNIST dataset, we compute a single correlation coefficient matrix, while it is clear that the matrix is highly class dependent. Further work will investigate whether prior clustering results in better performance, particularly in datasets of non-imaging origin.

## 5    Code and Data Availability

The code developed as part of this study is available at https://github.com/bariskanber/sdic. The datasets used are publicly available.